  \documentclass[10pt]{iopart}
\usepackage{iopams}	
\usepackage{epsfig,graphics}	
\begin{document}
 \newcommand{\centerfig}[2]{\centerline{\dofig{#1}{#2}}}
 \newcommand{\dofig}[2]{\resizebox{#1}{!}{\includegraphics{#2}}}
\title[Random solids and random solidification]{Random 
solids and random solidification:  What can be learned 
by exploring systems obeying permanent random 
constraints?\footnote{Presented at a workshop entitled
{\sl Unifying Concepts in Glass Physics\/}, 
held at the International Centre for Theoretical Physics, 
Trieste, Italy (September 15-18, 1999). 
To appear in a special issue of 
{\sl Journal of Physics~C\/}.}}

\author{Paul M Goldbart}

\address{Department of Physics, 
University of Illinois at Urbana-Champaign, \\
1110 West Green Street, Urbana, Illinois 61801-3080, U.S.A. \\
goldbart@uiuc.edu \\
http:$/\!/$www.pws.uiuc.edu/$\sim$goldbart \\ 
October 5, 1999
}
\begin{abstract}
In many interesting physical settings, such as the vulcanization of
rubber, the introduction of permanent random constraints between the
constituents of a homogeneous fluid can cause a phase transition to a
random solid state.  
In this random solid state, particles are permanently but randomly 
localized in space, and a rigidity to shear deformations emerges.  
Owing to the permanence of the random constraints, this phase transition 
is an equilibrium transition, which confers on it a simplicity (at least 
relative to the conventional glass transition) in the sense that it is 
amenable to established techniques of equilibrium statistical mechanics.  
In this Paper I shall review recent developments in the theory of random 
solidification for systems obeying permanent random constraints, with the 
aim of bringing to the fore the similarities and differences between such 
systems and those exhibiting the conventional glass transition.  I shall 
also report new results, obtained in collaboration with Weiqun Peng, on 
equilibrium correlations and susceptibilities that signal the approach 
of the random solidification transition, discussing the physical 
interpretation and values of these quantities both at the Gaussian level 
of approximation and, via a renormalization-group approach, beyond. 
\end{abstract}
%
%
%

\maketitle

\section{Introduction}
\label{SEC:Intro}
My aim in this article is to give a brief overview of recent developments 
in the theory of random solidification for systems obeying permanent random 
constraints.  Along the way, I hope to bring to the fore the similarities 
and differences between such systems and those exhibiting the conventional 
glass transition.   I shall not dwell on detailed technical matters, as they 
can be found in a number of articles which I will cite and which are 
readily available~\cite{REF:SeeThese}.  (Nor shall I attempt to provide 
a complete set of references.)\thinspace\ Instead, I shall focus on what I 
regard as essential matters of principle. 

The classic example of a system that undergoes random solidification 
in response to the imposition of a sufficient density of permanent 
random constraints on the motion of its constituents is vulcanized 
rubber.  Prior to vulcanization, the system
consists of a more or less viscous fluid of flexible macromolecules. 
Vulcanization amounts to the imposition of permanent covalent chemical 
bonds between randomly chosen atoms on the macromolecules. 
Both the chemical bonds defining the macromolecules and the bonds 
introduced by vulcanization will be regarded as permanent.  Although, of 
course, no chemical bonds are truly permanent, this means that we shall 
regard the breaking of these bonds to be extremely rare on the timescale 
needed for the unconstrained freedoms to equilibrate.  This separation 
of timescales provides the window necessary for the applicability of 
equilibrium statistical mechanics to the unconstrained freedoms, 
subject to the vulcanization constraints which play the role of 
quenched random variables. 

A second example is provided by chemical gels, such as silica gel, 
in which low-molecular-weight objects (rather than macromolecules) 
are permanently bonded together at random (e.g.~by a 
condensation/elimination reaction) so as to build up a giant random 
molecule.  It is useful to regard chemical gels, too, as systems 
that undergo random solidification in response to the introduction 
of a sufficient density of permanent random constraints (i.e.~random 
covalent bonds). 

What do we mean when we say: Random permanent constraints lead to random 
solidification?  We mean that the system undergoes a phase transition 
(as it happens, continuous), as the density of constraints is increased 
beyond some critical value, with the following characteristics. At 
subcritical  densities the system is fluid in the following two senses: 
(i)~it does not respond to an applied zero-frequency shear  strain by
developing a zero-frequency shear stress; and (ii)~none of the particles in
the system is localized and instead,  given sufficient time, all  wander
throughout the (essentially infinite) container.  On the other  hand, at
supercritical densities of constraints the system is a solid  in both
senses: (i)~it responds to a applied zero-frequency  shear strain by 
developing a zero-frequency shear stress; and (ii)~at  least a fraction  of
the particles in the system are localized, so that instead of wandering 
throughout the container they remain in the vicinity of their mean 
positions, from which they make thermally-driven excursions of only finite 
spatial extent.  The fundamental competition leading to the transition 
is one between translational entropy, which favors delocalization, and 
crosslinking, which favors localization.  Note that the transition is 
driven by crosslink density, not temperature.

As for the randomness of the emergent solid, this means that although 
translational symmetry is spontaneously broken---at least some particles 
acquiring random mean positions about which they undergo thermal 
motion---it is broken randomly, in the sense that there is no 
crystalline long-range order to the collection of mean positions: Fourier 
analysis, as we shall see more concretely in Sec.~\ref{SEC:DRS}, 
detects no long-range periodicity to the mean positions.  Furthermore, 
there is a second level of randomness to the emergent solid.  Not only 
are the mean particle positions distributed at random throughout the 
container but, also, every localized particle experiences a distinct 
environment.  This shows up, e.g., via a statistical continuum of 
r.m.s.~displacements  of the localized particles (measured from their 
mean positions), i.e., a statistical distribution of localization lengths.  

There is a subtlety here concerning the notion of localization that is 
worth commenting on.  At temperatures below its melting temperature, the 
equilibrium state of, say, the element Cu is a crystalline solid,  in
the sense that an infinitesimal stress produces an infinitesimal strain 
(rather than strain {\it rate\/}, as it would in a fluid).  However, the 
Cu atoms are not localized: vacancy motion causes them to diffuse, albeit 
slowly, throughout the crystal.  
In contrast, the  permanence of the chemical bonds in
vulcanized macromolecular systems, both intra- and inter-molecular, at
least on the timescale of envisaged experiments, allows for true
localization.  In this sense,  vulcanized matter forms solids that are
simpler than \lq\lq simple\rq\rq\  solids, such as crystalline Cu. 
\section{Connections with glasses: freezing in of correlations}
\label{SEC:CWG} 
Now, the title of the conference for which this Paper was prepared is 
{\it Unifying Concepts in Glass Physics\/}, so I would like to make 
some general remarks about the relationship between vulcanized 
systems and the kinds of systems that are conventionally referred to 
as glassy.  For reasons that I hope will become apparent, I believe 
that it may be profitable to regard vulcanized systems as model 
glasses (see Ref.~\cite{REF:SGRFbook}).  

Presumably, {\it conventional\/} glassy regimes emerge when certain
structural  correlations, characteristic of a liquid state and involving
the positions  of the particles, of are unable to relax on the timescale of
an experiment,  and are therefore frozen in.  (Note that I am sidestepping
the fundamental  question of whether or not such correlations can be frozen
in on infinite  timescales.)\thinspace\   One of the central obstacles to
the development of a sound theory of this  phenomenon of glassiness is
that, at least {\it prima facie\/}, it appears  intrinsically dynamical
(and perhaps is so at its very core).  If so, this  would confer on the
subject a level of technical complexity, relative to  issues that are
amenable to the techniques of equilibrium statistical  mechanics for
systems possessing {\it extrinsic\/} (rather than  {\it intrinsic\/}, i.e.,
spontaneously generated) quenched random  constraints.  The origin of the
relative simplicity in the latter case 
is that one is presented with a wide (and
evident) separation between the  (short) timescale for the equilibration of
unconstrained freedoms and the  (long) timescale for the constraints to
break.  This separation yields a  window of times in which equilibrium
statistical mechanics is obviously  applicable, and allows one to
confidently finesse the task of statistical  physics on timescales
comparable to any that are intrinsic to the system,  which is, of
necessity, a matter of dynamics.

Extrinsic quenched random constraints, such as those introduced by the 
vulcanization process, have {\it precisely\/} the effect of freezing 
in certain structural correlations characteristic of a liquid state.  
However, not being intrinsic to the liquid, extrinsic constraints have 
an independence from it that allows one to tune the density of them and, 
hence, to {\it tune the state of the system in a controllable fashion 
right through the random solidification transition\/}.  In this sense, 
with vulcanized matter one has access to a critical \lq\lq glass\rq\rq\ 
transition that one does not have with intrinsically glassy systems.  For 
the latter, the reliance on self-generated constraints necessarily 
creates a (presumably highly significant) \lq\lq feedback loop\rq\rq\ 
between the correlations that are frozen in and the consequent state 
of the system.  It does not seem unreasonable to suppose that this 
feedback effect is what causes the critical random solidification 
transition to be pre-empted by the conventional glass transition. 

This point of view, which was elaborated in Ref.~\cite{REF:SGRFbook}, 
served as the main motivation for an exploratory approach to structural 
glasses begun in Ref.~\cite{REF:SMCRN:mgt} and developed in considerable 
detail in Ref.~\cite{REF:SMRN:sgs}.  In this approach, one considers the 
equilibrium statistical mechanics of 
a network built from low-molecular-weight molecules 
connected at random by permanent covalent bonds.  The 
formation of silica gel networks via a polycondensation/elimination 
reaction provides a concrete example of such networks; see, e.g., 
Ref.~\cite{REF:Gauthier}.   The essence of this approach is to 
identify correlations of the liquid state (i.e.~particles separated 
by a bond distance), to freeze in some fraction of these correlations 
(as quenched random constraints), and to consider how the system 
responds.  By this scheme, we are \lq\lq pushing off to infinity\rq\rq\ 
the time scale for the relaxation of these correlations \lq\lq by 
hand\rlap,\rq\rq\ and thereby establishing a setting amenable to the 
techniques of equilibrium statistical mechanics. 
This is a perfectly reasonable strategy for systems such as silica gel, 
in which there is a clear separation of timescales; for conventional 
glasses the strategy represents an idealization, and the question 
remains: To what extent is it a useful one? 

What emerges from this approach to glassy systems?  
Put briefly (see Ref.~\cite{REF:SMRN:sgs} 
for details), one finds a phase transition from a liquid to a random solid 
state characterized by a rich order parameter, which 
encodes information not only about the positional localization of the 
particles in the system but also about the consequent orientational 
localization of the bonds that connect the particles (as well as thermal 
correlations between particle positions and bond orientations). 

I think it is worth emphasizing that the approach to random solidification
outlined here  is in fact rather straightforward (at least in spirit, if
not practice): A semi-microscopic  model leads to a field theoretic
representation that is amenable to a (rich and informative) saddle-point
approximation and, as we shall see, systematic improvements via the
application of renormalization-group ideas. 

\section{Detecting random solids: An order parameter}
\label{SEC:DRS} 
Let us briefly explore an order parameter capable of detecting the  random
solid state.  For detailed discussions, see 
Refs.~\cite{REF:RCMSreview,REF:SGRFbook,REF:Rigidity}. Consider a
collection of $N$ particles, labelled $j=1,\ldots,N$ and  having
$d$-dimensional positions $\{{\bf R}_{j}\}_{j=1}^{N}$.   (The slight 
elaboration required to handle macromolecular freedoms rather than  point
particles is not of importance to the present discussion.)\thinspace\  The
basic element to focus on is the Fourier transform of the  probability
density  of finding the particle to be at some position ${\bf r}$ in the
volume $V$, viz., 
\begin{equation}
\int_{V}\rmd^{d}r\,
\big\langle\delta^{(d)}({\bf r}-{\bf R}_{j})\big\rangle\,
\exp\rmi{\bf k}\cdot{\bf r}
=\big\langle\exp\rmi{\bf k}\cdot{\bf R}_{j}\big\rangle, 
\end{equation}
where $\langle\cdots\rangle$ denotes an equilibrium expectation value 
of a randomly constrained system, perhaps in a broken symmetry state. 
If particle $j$ is {\it delocalized\/} then, by translational invariance, 
$\big\langle\exp\rmi{\bf k}\cdot{\bf R}_{j}\big\rangle=
\delta_{{\bf k},{\bf 0}}$. 
On the other hand, if it is {\it localized\/} then a good model to bear 
in mind for later use is
\begin{equation}
\big\langle\exp\rmi{\bf k}\cdot{\bf R}_{j}\big\rangle\approx
\exp\left(\rmi{\bf k}\cdot\langle{\bf R}_{j}\rangle\right)\, 
\exp\left(-\xi_{j}^{2}k^{2}/2\right), 
\label{EQ:model}
\end{equation}
i.e., the Fourier transform of a Gaussian distribution for the particle 
position, characterized by the mean position $\langle{\bf R}_{j}\rangle$ 
and the r.m.s.~displacement (i.e.~{\it localization length\/}) $\xi_{j}$.  
For future use, it is 
worth noting the distinction between the ${\bf k}={\bf 0}$ value and 
the ${\bf k}\to{\bf 0}$ limit of 
$\vert\langle\exp\rmi{\bf k}\cdot{\bf R}_{j}\rangle\vert$. 
Of course, regardless of whether or not particle $j$ is localized one 
obtains unity precisely at ${\bf k}={\bf 0}$; by contrast, however, 
in the limit one finds zero if particle $j$ is delocalized but unity 
if particle $j$ is localized.  

The first construct a statistical mechanician might examine is the 
average of 
$\big\langle\exp\rmi{\bf k}\cdot{\bf R}_{j}\big\rangle$ 
over the particles in the system: 
\begin{equation}
{1\over{N}}\sum_{j=1}^{N}
\big\langle\exp\rmi{\bf k}\cdot{\bf R}_{j}\big\rangle.
\label{EQ:onerep}
\end{equation}
Not surprisingly, if all particles are delocalized then this quantity
takes  the value 
$\delta_{{\bf k},{\bf 0}}$.  
However, if some fraction of the $N$  particles are localized but their
mean positions 
$\langle{\bf R}_{j}\rangle$  
are random then, for no ${\bf k}$ except 
${\bf 0}$ do the
contributions add  constructively, so that the sum
continues to take the  value 
$\delta_{{\bf k},{\bf 0}}$.   
Thus the entity~(\ref{EQ:onerep}) fails to  distinguish between the
delocalized liquid and the randomly localized solid, and does so in the
much same way that the magnetization density fails to distinguish between, 
e.g., the spin glass and paramagnetic states. 

So, what's the remedy?  As with the spin glass case, to avoid destructive 
interference one considers more than one equilibrium expectation value 
under the average over particles (and then disorder-averages; 
see Sec.~\ref{SEC:DXL}): 
\begin{equation}
\bigg[
{1\over{N}}\sum_{j=1}^{N}
\prod_{\alpha=0}^{g}
\big\langle
\exp\rmi{\bf k}^{\alpha}\cdot{\bf R}_{j}
\big\rangle
\bigg], 
\label{EQ:morrep}
\end{equation}
where $[\cdots]$ denotes disorder-averaging. If all $N$ particles are 
delocalized then this entity only fails to vanish for the trivial case 
of all wave vectors 
$\{{\bf k}^{\alpha}\}_{\alpha=0}^{g}$ 
vanishing.  However, if some fraction $q$ of the particles are localized 
with random mean positions then the entity is nonzero 
whenever $\sum_{\alpha=0}^{g}{\bf k}^{\alpha}$ 
vanishes, i.e., the random solidification transition is associated with 
the change in form of~(\ref{EQ:morrep}) from being a \lq\lq bump\rq\rq\ 
concentrated at the origin $\{{\bf k}^{\alpha}={\bf 0}\}_{\alpha=0}^{g}$ 
of the replicated wave vector space, 
to being a \lq\lq fin\rq\rq\ concentrated on the hypersurface  
$\sum_{\alpha=0}^{g}{\bf k}^{\alpha}={\bf 0}$.  Thus we see 
that~(\ref{EQ:morrep}) serves as an order parameter for the liquid to 
amorphous solid phase transition.  Moreover, the limit 
$\{{\bf k}^{\alpha}\to{\bf 0}\}_{\alpha=0}^{g}$ 
(with $\sum_{\alpha=0}^{g}{\bf k}^{\alpha}={\bf 0})$ 
determines the {\it fraction\/} of localized particles, and the shape 
of the fin determines the {\it distribution\/} of localization lengths.  
For example, if the localized fraction is localized according to 
Eq.~(\ref{EQ:model}) then the order parameter~(\ref{EQ:morrep}) 
would take the form
\begin{equation}
(1-q)\prod_{\alpha=0}^{g}\delta_{{\bf k}^{\alpha},{\bf 0}}
	+q\,\delta_{\Sigma_{\alpha=0}^{g}{\bf k}^{\alpha},{\bf 0}}\,
\int\rmd\xi\,p(\xi)\,
\exp\left(-{\xi^{2}\over{2}}
\sum_{\alpha=0}^{g}\vert{\bf k}^{\alpha}\vert^{2}\right), 
\label{EQ:ansatz}
\end{equation}
where 
$p(\xi)\equiv
\big[(qN)^{-1}\sum_{j\,\,{\rm loc.}}\delta^{(1)}(\xi-\xi_{j})\big]$ 
is the statistical distribution of the random localization lengths 
of the localized particles.  

We note, in passing, that under Fourier 
transformation to real space [via 
the operation 
$V^{-1}\sum_{{\bf k}^{0}}
\exp\big(-i{\bf k}^{0}\cdot{\bf r}^{0}\big)
\cdots
V^{-1}\sum_{{\bf k}^{g}}
\exp\big(-i{\bf k}^{g}\cdot{\bf r}^{g}\big)\cdots$]
this order parameter becomes
\begin{equation}
\fl
{1-q\over{V^{g+1}}}
+{q\over{(2\pi)^{d(g+1)/2}}}
\int\rmd\xi\,p(\xi)\,\xi^{-d(g+1)}
\int_{V}
{\rmd^{d}\rho\over{V}}
\exp\left(-{1\over{2\xi^{2}}}\sum_{\alpha=0}^{g}
\vert{\bf r}^{\alpha}-\brho\vert^{2}
\right). 
\label{EQ:MBS}
\end{equation}
If we regard the $g+1$ replicas $\{{\bf r}^{\alpha}\}_{\alpha=0}^{g}$ of 
the particle position as the positions of the $g+1$ \lq\lq atoms\rq\rq\ of 
a \lq\lq molecule\rq\rq\ then this joint probability distribution 
describes a {\it molecular bound state in replica space\/}.  That it is
bound follows from localization; invariance under common translations of
the atoms follows from the randomness of the mean locations of the
localized  particles; the shape of the bound state follows from the 
distribution of localization lengths.  
\section{Characteristics of the random solid state}
\label{SEC:CRSS} 
So far, we have introduced an order parameter capable of detecting 
and diagnosing the random solid state.  We now turn to the question 
of computing it within various frameworks and for various model settings.
\subsection{Semi-microscopic replica theory for vulcanized macromolecules}
\label{SEC:SMRT} 
The most direct approach to the computation of the order parameter 
arises in the setting of randomly crosslinked macromolecular systems, and 
builds upon the formulation of the statistical mechanics of such systems 
established in the beautiful work of Deam and Edwards~\cite{REF:DE:TRE}.  
In the present article I shall just sketch the strategy, and 
encourage the reader to turn to  Refs.~\cite{REF:RCMSreview,REF:UOAST} 
for technical details. 
\subsubsection{Partition function}
\label{SEC:Partition} 
We begin with the partition function for a system of $N$ identical, 
interacting, flexible, uncrosslinked (and hence non-random) 
macromolecules, this partition function describing the fluid 
(i.e.~melt or solution) of uncrosslinked macromolecules.  (We view 
these macromolecules at the semi-microscopic level, which means that 
we ignore details of their chemical constitution and, instead, 
consider featureless strands of matter.)
\subsubsection{Crosslinks as quenched random constraints}
\label{SEC:XQRV} 
Now, how do we incorporate into this semi-microscopic description 
the effects of random crosslinking?  In three spatial dimensions, 
at least, crosslinking has two distinct effects, which 
we may choose to call {\it holonomic\/} and {\it anholonomic\/}.  
The holonomic effect 
of each crosslink serves to {\it identify\/} the positions of two points 
on the macromolecules, chosen at random, i.e., to introduce a random 
constraint that identifies two positional freedoms that formerly were 
kinematically independent.  The anholonomic effect serves to select a 
specific topological structure for the network: Given 
the collection of macromolecules and the collection of holonomic 
constraints, how are the macromolecules \lq\lq woven\rq\rq\ into a 
$d$-dimensional network?  The {\it quenched random information\/} 
is then: (i)~the catalogue of pairs of points that are identified 
by the crosslinks; and (ii)~the specific topology of the macromolecular 
strands subject to the identification of the pairs of crosslinked points. 
As we shall discuss shortly, the holonomic aspect of crosslinking can 
readily be incorporated; however, we know of no scheme capable of 
accounting for the anholonomic aspect.  Thus, in practice we shall 
treat the holonomic aspect as quenched random information, 
but shall treat fluctuations between distinct network topologies 
as annealed variables.  This is not totally indefensible. 
First, the transition regime, which is the regime of interest to us 
here, is characterized by rather light crosslinking---of order 
one crosslink per macromolecule---so the effects of topology might 
reasonably be expected to be weak.  
Second, if one imagines coarse-graining one's view of the system 
then the distinction between holonomic and anholonomic constraints 
tends to blur, with knots and crosslinks having rather similar effects.

The next step is to specify, at random, a catalogue of pairs of 
points that are to be identified by the crosslinks, i.e., to specify 
the quenched random information that determines the holonomic aspect 
of the constraints.  (Shortly, we shall discuss the issue of how to 
construct a reasonable model distribution for this quenched random 
information.)\thinspace\ We use this information to remove from the 
sum over configurations (which constitutes the partition function 
for the uncrosslinked system of macromolecules) all configurations 
that fail to satisfy the (holonomic aspect of the) constraints.  
In practice this removal is accomplished by a suitable product of 
Dirac delta functions, which is zero for configurations that do not 
obey the constraints.
The resulting partition function describes a specific realization 
of the randomly crosslinked system and, just as the partition 
function for a spin glass depends on the quenched random interactions, 
this partition function depends on the quenched random information 
describing the constraints.  As such, it is impractical to handle 
directly and, for the usual reasons, one averages its logarithm, 
essentially the free energy of the randomly constrained system, 
over some distribution of the quenched random information. 
\subsubsection{Distribution of crosslinks}
\label{SEC:DXL} 
How might one ascribe a statistical weight to a specific realization 
of random crosslinks?  One strategy, which is due to Deam and Edwards 
and which I regard as extremely elegant~\cite{REF:DE:TRE}, is to 
imagine the following experimental procedure.  
Take the uncrosslinked liquid in equilibrium.  Stop time.  
(Then the likelihood of finding any configuration will be proportional 
to its Boltzmann weight.)\thinspace\  Examine the configuration that 
you have, and identify points of near-contact between macromolecules. 
Independently for each near-contact, either do or do not introduce a 
crosslinking constraint with some probability.  The distribution of 
crosslinks thus constructed has the virtue of being determined by the 
equilibrium state of the uncrosslinked liquid, together with a single 
number, the crosslinking probability, the latter governing the (mean 
number) of crosslinks introduced and ultimately playing the role of 
the control parameter for the phase transition from the liquid to the 
random solid state.  

In fact, by choosing this Deam-Edwards crosslink distribution one is 
conferring upon the system an additional---and highly 
convenient---symmetry.  The origin of this symmetry is the fact, 
which will become obvious to the reader after a moment's reflection, 
that the Deam-Edwards crosslink probability distribution is itself 
proportional to the partition function of the randomly crosslinked 
system.  Perhaps this observation is not of much comfort at 
the present stage, inasmuch as it only tells us that the theory 
contains one, rather than two, entities that we have not yet 
managed to compute.  But at least it is one and not two!  And 
shortly we shall see that this symmetry provides the reason why 
the replica theory turns out to have $(n+1)$-fold, rather than 
the usual $n$-fold, permutation symmetry. 

It should be stressed that there is much more to this choice of 
crosslink distribution than the enhanced symmetry it confers.  
Even more significantly, being based on a physical vulcanization 
process it ensures that appreciable statistical weight is only 
given to those realizations of the crosslinking that the 
macromolecules can accommodate, and thus it leads to reasonable, 
space filling, statistically homogeneous (random solid) states. 
\subsubsection{Replica statistical mechanics; replica field theory}
\label{SEC:RSMRFT} 
Taking stock of the situation, we see that we have a theory with: 
(i)~annealed variables (the macromolecular coordinates), 
(ii)~quenched variables (the number and specification of the crosslinks), 
(iii)~a random partition function (containing random {\it constraints\/} 
on \lq\lq street level\rq\rq\ rather than random {\it interactions\/} 
in the exponent of the Boltzmann weight), and 
(iv)~a distribution for the quenched randomness based on a physical 
model for the vulcanization process.  In the context of other random 
systems the distribution might, e.g., be a Gaussian distribution of 
exchange interactions (spin glasses) or of random impurity potentials 
(electronic transport). 

Proceeding in the familiar way, we use the replica technique to compute 
the disorder-average of the logarithm of the partition function.  
However, in our setting of randomly constrained systems what 
emerges from this procedure may be quite unfamiliar.  First, 
owing to the \lq\lq street level\rq\rq\ location of the constraint 
delta functions, together with the random {\it number\/} of them, 
averaging over the quenched random variables yields a term in the 
exponent of the (effective, pure) Boltzmann weight featuring a 
{\it product over replicas\/}, in contrast with the more familiar 
{\it pairwise coupling\/} of replicas that commonly results from 
random {\it interactions\/} (cf.~exchange interactions in spin 
glasses or random impurity potentials in electronic transport).   
This term has the effect of causing all pairs 
of segments of the replicated macromolecules to attract one another 
in replicated space.  Second, as mentioned above, the equilibrium 
average involved in the construction of the crosslink distribution 
provides an additional replica, so that we end up with the $n\to 0$ 
limit of a theory involving $n+1$, rather than the usual $n$, 
replicas.  

Now, how can we address the resulting pure, replica theory?   We may, as is
often done  for spin glasses, apply a Hubbard-Stratonovich decoupling 
transformation, which leads to a field theory representation.  (The 
virtues of this procedure are manifold: it suggests the physically 
appropriate collective coordinates; it allows us to attack the  problem
using calculus; and it is exact, and thereby provides us  with a framework
for going beyond mean-field theory.)\thinspace\  In the present context the
necessary field turns out to be  complex-valued and to live on $(n+1)$-fold
replicated space, 
$\Omega({\bf r}^{0},{\bf r}^{1},\ldots,{\bf r}^{n})$; it is precisely 
the Fourier transform of the order parameter introduced and  motivated on
physical grounds in Sec.~\ref{SEC:DRS}.   A standard linear-stability 
analysis of this field theory reveals that there is indeed a  phase
transition, as the density of crosslinks is increased.  Moreover, provided
the underlying fluid consists of adequately  repulsive macromolecules, the
stability analysis indicates that  the instability is precisely of the
physically anticipated form:  the liquid state becomes unstable and the
instability is in the  direction of the random solid state.

But how does the instability actually get resolved?  In other words, 
what is the precise form of the state that replaces it?  Let us begin 
by addressing this question at the level of mean-field theory.  
(Going beyond mean-field theory will be discussed below, in 
Sec.~\ref{SEC:Critical}.)\thinspace\ 
This natural starting point can be accomplished 
in several equivalent ways.  One may treat the field theory mentioned 
above at the saddle-point level.  Or one can take more direct route 
of replacing fluctuating collective coordinates by their average 
values plus departures, and then linearizing in the departures.
Whatever the scheme, one ends up with a self-consistent equation 
for the order parameter: a functional equation for the field 
$\Omega({\bf r}^{0},{\bf r}^{1},\ldots,{\bf r}^{n})$. 

\begin{figure}
\begin{center}
\begin{tabular}{cc}
\psfig{figure=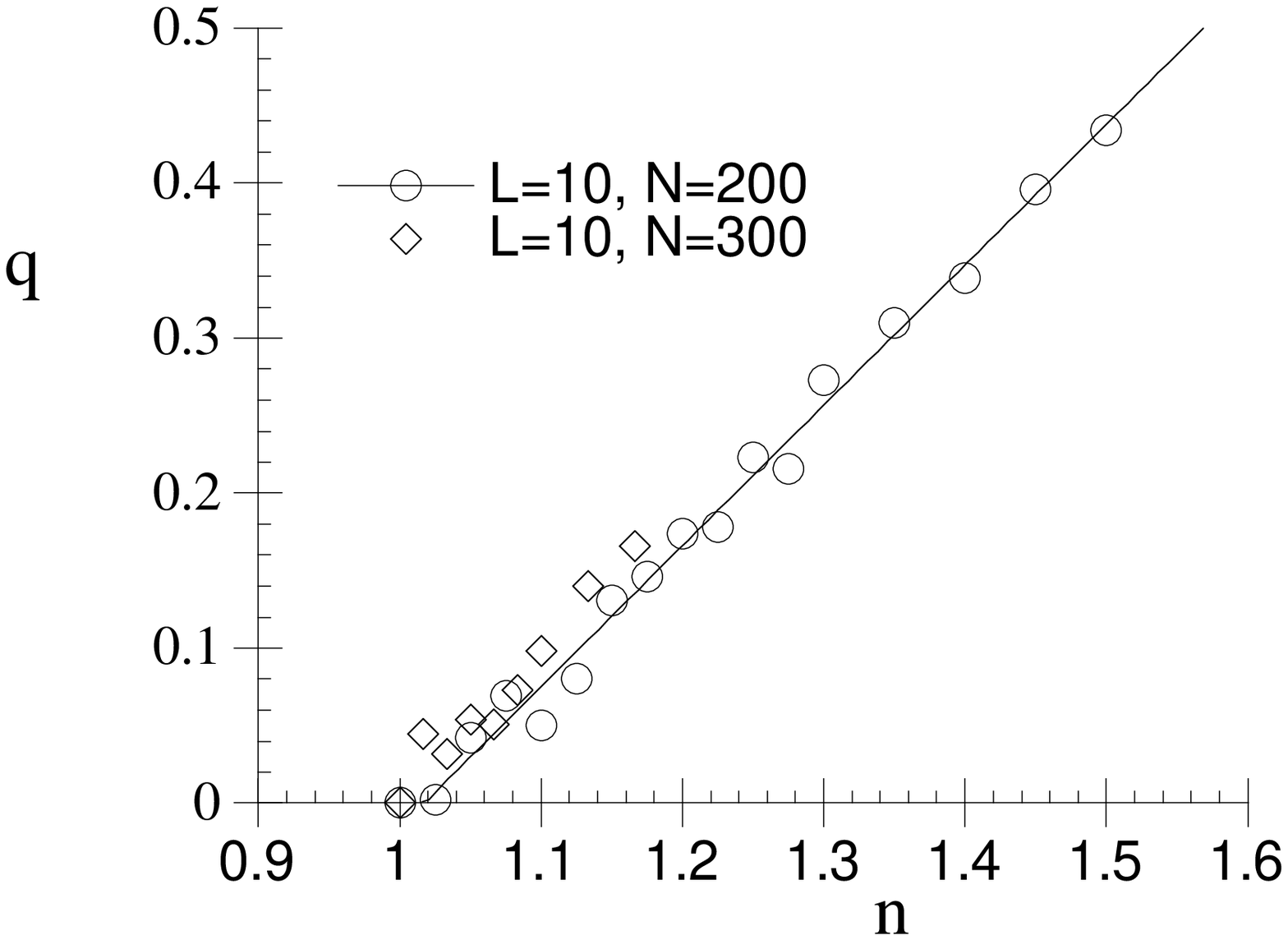,width=2.95in} & 
\psfig{figure=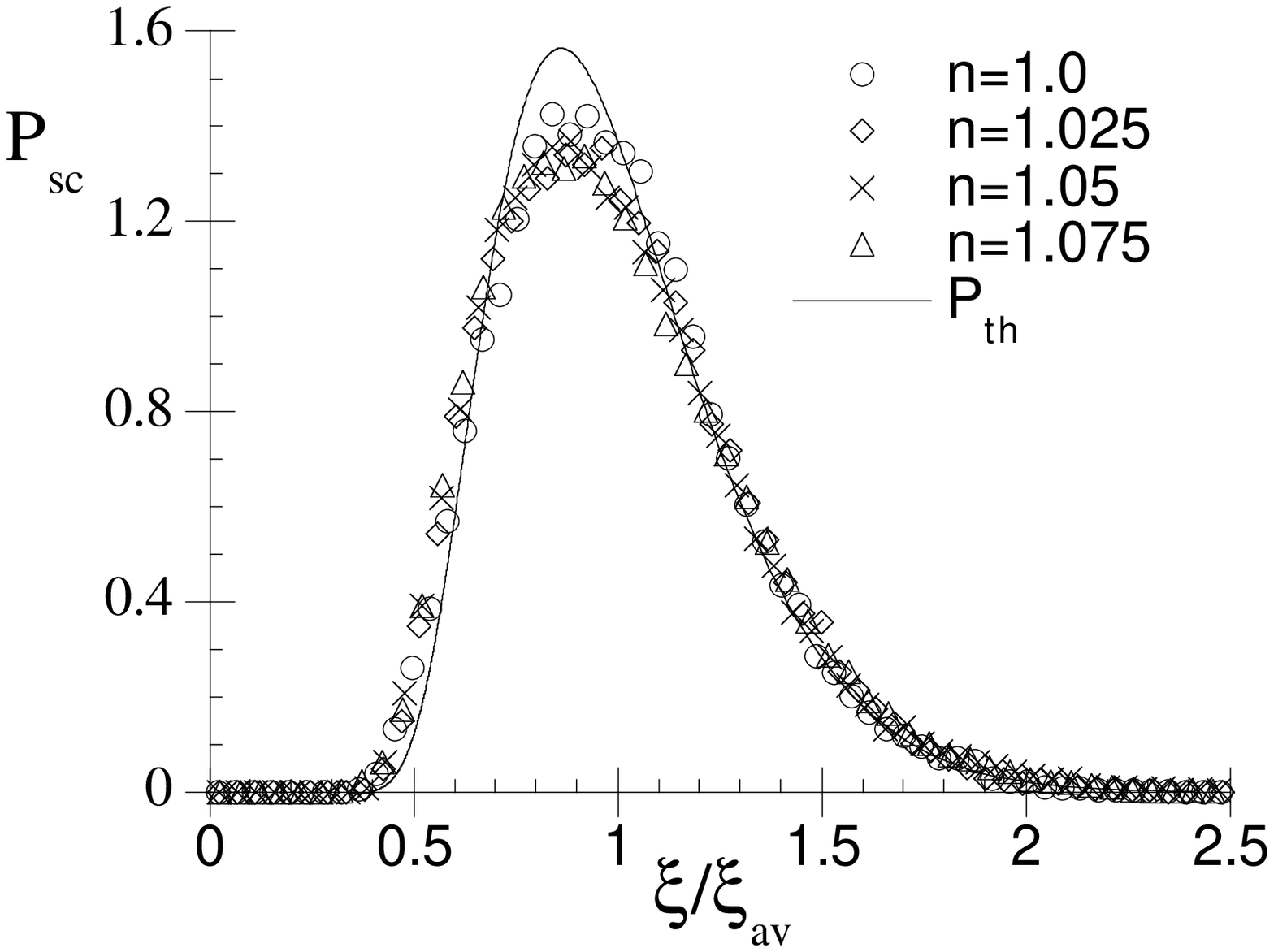,width=2.95in} 
\end{tabular}
\end{center}
\caption{
Results from molecular dynamics simulations by 
Barsky and Plischke (1997, unpublished). 
{\it Left\/}: 
Localized fraction $q$ versus number of 
crosslinks per macromolecule $n$.
$L$ is the number of monomers per macromolecule; 
$N$ is the number of macromolecules in the system.  
The straight line is a fit to the $N=200$ data.
Note the continuous phase transition at $n=1$ 
and the linear variation of $q$ with $n$, both 
consistent with mean-field theory.
{\it Right\/}:
Distribution $P_{\rm s}$ of localization lengths $\xi$ (symbols), 
scaled with the sample-average localization lengths $\xi_{\rm av}$ 
Note the collapse of the data on to a universal scaling distribution, 
and the quantitative agreement with the mean-field prediction (solid line). 
The number of segments per macromolecule was $10$; 
the number of macromolecules was $200$.
\label{FIG:composite}}
\end{figure}

In general, this equation is highly complicated, even to write down 
let alone solve.  However, in the random solid state, but near to the 
transition, the smallness of the fraction of localized particles 
provides a simplification, allowing us to discard all but the 
quadratic nonlinearity in the self-consistent equation for the order 
parameter, which then reads 
\begin{equation}
0=
2\Big(-a\epsilon+\frac{b}{2}\vert{\hat{k}}\vert^2\Big)
\Omega({\hat{k}} )
-3c\overline{\sum\limits_{\hat{k}_1\hat{k}_2}}
\Omega({\hat{k}_1})\,
\Omega({\hat{k}_2})\,
\delta_{{\hat{k}_1}+{\hat{k}_2},{\hat{k}}}\,.
\label{EQ:SCE}
\end{equation}
Here, $a$, $b$ and $c$ are model-dependent coefficients, 
$\epsilon$ measures the excess crosslink density (beyond the 
mean-field critical value of zero), 
and 
$\Omega({\bf k}^{0},{\bf k}^{1},\ldots,{\bf k}^{n})$ 
is the Fourier transform of 
$\Omega({\bf r}^{0},{\bf r}^{1},\ldots,{\bf r}^{n})$. 
To ease the notation we have taken to writing 
$\hat{k}$ for 
$\{{\bf k}^{0},{\bf k}^{1},\ldots,{\bf k}^{n}\}$ 
so that  
$\Omega({\bf k}^{0},{\bf k}^{1},\ldots,{\bf k}^{n})$ 
becomes 
$\Omega(\hat{k})$ 
and 
$\sum_{\alpha=0}^{n}{\bf k}^{\alpha}\cdot{\bf k}^{\alpha}$
becomes 
$\vert{\hat{k}}\vert^2$. 
The overbar on the summations indicates that a certain class of 
terms is to be omitted from the summations: the reason for this 
(vital) restriction will be discussed shortly.  It is not 
difficult to verify that this equation is {\it exactly\/} solved 
by the function given in Eq.~(\ref{EQ:ansatz}), provided one 
chooses $g=n$ (with $n\to 0$), and (up to simple rescalings 
involving $a$, $b$ and $c$) the fraction of localized particles $q$ 
and the normalized distribution of localization lengths $p(\cdot)$ 
to be given by 
\begin{eqnarray}
q&=&2\epsilon/3,
\\
p(\xi)&=&(4/\epsilon\xi^{3})\,\pi (2/\epsilon\xi^{2}), 
\label{EQ:TwoEqs}
\end{eqnarray}
where the scaling function $\pi(\cdot)$ obeys the simple nonlinear 
integro-differential equation
\begin{equation}
{\theta^{2}\over{2}}{d\pi\over{d\theta}}=
(1-\theta)\,\pi-
\int_{0}^{\theta}\rmd\theta^{\prime}\,
\pi(\theta^{\prime})\,
\pi(\theta-\theta^{\prime}), 
\label{EQ:NLE}
\end{equation}
the solution of which gives rise to the parameter-free prediction 
shown as the full line in Fig.~\ref{FIG:composite} (right).

So, what have we found?  At this stage we have found that a mean-field 
treatment of a specific semi-microscopic model yields the following 
results: 
(i)~For crosslink densities smaller than a certain critical value
(of roughly one crosslink per macromolecule) the equilibrium state 
of the system is a liquid, with all particles (in the context of 
macromolecules, monomers) being delocalized.  
(ii)~At the critical crosslink density there is a continuous
thermodynamic phase transition to a random solid state 
characterized by the presence of random static density fluctuations.
(iii)~In this state, at least a fraction of the particles are 
localized near random positions about which they thermally 
fluctuate with random localization lengths. 
(iv)~The fraction of localized particles grows linearly with the  excess
crosslink density,  consistent with the mean-field theory of percolation. 
As the  transition is approached, the characteristic localization length 
diverges as the inverse square root of the excess crosslink density. 
(v)~When scaled by the mean value, the statistical distribution of
localization lengths is universal for all near-critical crosslink
densities, the form of this scaled distribution being uniquely determined
by a the integro-differential equation~(\ref{EQ:NLE}). 

As one can see from Fig.~\ref{FIG:composite}, this picture has 
been rather well confirmed in molecular dynamics computer 
simulations of three-dimensional systems of randomly crosslinked, 
interacting macromolecules, undertaken at Simon Fraser 
University by Barsky and Plischke~\cite{REF:SJB_MP_a,REF:SJB_MP_b}. 
\subsection{Universality; simulations; a Landau theory}
\label{SEC:USLT} 
By repeating the strategy outlined above for several other systems 
that undergo random solidification, including endlinked (flexible 
and semi-flexible) macromolecular systems~\cite{REF:Endlinking}, chemically 
gelled low-molecular-weight systems~\cite{REF:SMCRN:mgt,REF:SMRN:sgs}, 
and even cross{\-}linked manifolds~\cite{REF:Roos}, we have learned that the 
critical properties (i.e.~the exponent for the localized fraction 
and the scaling function for the distribution of localization lengths) 
are {\it universal\/} over a broad class of systems, at least at the 
mean-field level.  (The extension of this universality beyond mean-field 
theory will be discussed in Sec.~\ref{SEC:Critical}.)\thinspace\ 
Further evidence for this universality has come from the extensive 
sequence of computer simulations mentioned above. 

This universality at the mean-field level can be understood
from the perspective of a model-independent 
Landau approach~\cite{REF:UOAST}. 
Suppose we take for granted the idea that the order parameter 
is a complex-valued field on $(n+1)$-fold replicated Fourier 
space: $\Omega(\hat{k})$. 
Let us assume that near to the transition the fraction 
of localized particles is small and that the localization length 
of the localized particles is large.
Let us also assume that macroscopic density fluctuations are 
suppressed by inter-particle interactions, and therefore 
do not fluctuate critically at the random solidification transition.
As such fluctuations correspond to field configurations in 
which, e.g., $\Omega({\bf 0},{\bf k},{\bf 0},\ldots,{\bf 0})$ 
is nonzero, we should only allow equilibrium values of $\Omega$ 
that vanish whenever all but one (or all) argument-vectors vanish. 
Then (up to simple coefficients) we arrive at the following 
Landau free energy (per object---e.g.~macromolecule---being 
crosslinked): 
\begin{equation}
\fl
{\overline{\sum}}_{\hat{k}}
\big(-a\epsilon+\frac{b}{2}|\hat{k}|^2\big)
\big\vert\Omega({\hat{k}})\big\vert^{2}
-c\,{\overline{\sum}}_{{\hat{k}_1}{\hat{k}_2}{\hat{k}_3}}
\Omega({\hat{k}_1})\,
\Omega({\hat{k}_2})\,
\Omega({\hat{k}_3})\,
\delta_{{\hat{k}_1}+{\hat{k}_2}+{\hat{k}_3}, {\hat{0}}}\,.
\label{EQ:LFE}
\end{equation}
The innocuous-looking condition that $\Omega$ be zero whenever all but one 
(or all) argument-vectors vanish, alluded to just after Eq.~(\ref{EQ:SCE}),  
amounts to our incorporating into the theory the effects of interparticle 
repulsions, and is vital.  

As one can see from the quadratic 
term in the Landau theory, supercritical crosslinking causes the 
liquid state to become unstable.  However, the 
instability cannot be resolved as it would be, say, in the case of 
ferromagnetism with its homogeneous ferromagnetic state, by the 
\lq\lq condensation\rq\rq\ of (i.e.~acquisition of a non-zero 
value by) the homogeneous mode because there is no such mode in the 
theory.  Instead, non-zero wave vector modes acquire non-zero values, 
and do so in a delicate balance determined by the cubic interaction 
term, so as to form a kind of {\it domain wall in momentum space\/}.  
What is especially delicate is the scheme by which the system avoids 
condensing in the (forbidden) macroscopic density-sector.  By having 
condensation only of modes for which 
$\sum_{\alpha=0}^{n}{\bf k}^{\alpha}={\bf 0}$, 
the cubic interaction term cannot induce the condensation of 
macroscopic density-sector fluctuations.  One might say that the 
competing tendencies of the crosslinks (which cause attraction 
simultaneously in replica space) and the interparticle repulsion 
(which acts separately in each replica) frustrate one another, 
this frustration resolving itself via random solidification. 
Owing to its macroscopic translational invariance, this state 
is a condensation that manages to avoid the energy cost that 
macroscopic density-sector fluctuations would bring.  

The condition on the field is also vital from a symmetry standpoint, 
and embodies the notion that, provided interparticle interactions are 
included, the only symmetry of the theory associated with the mixing 
of the replicas is the {\it permutation\/} symmetry $S_{n+1}$.  The 
replica coupling arising from the disorder-averaging of the replicated 
crosslinking constraints yields a term that has the special feature of 
being invariant under the larger group ${\rm O}\left((n+1)d\right)$ of 
rotations that mix the (Cartesian components of the) replicas. However, 
this symmetry is explicitly broken down to a {\it permutation\/} 
symmetry by the interparticle interactions.  This reduced symmetry also 
plays vital roles the analysis of the local stability of the random 
solid state and the determination of the universality class of the phase 
transition in the context of a renormalization-group approach to it, 
as we shall see further in Sec.~\ref{SEC:Critical}. 

For specific values of the coefficients, this Landau theory is  what
emerges as the replica mean-field theory for  each of the semi-microscopic
models considered (except that the  theory is a bit more elaborate for the
gelation case).  It is therefore not surprising that it, too, is made
stationary  by the form given in Eq.~(\ref{EQ:NLE}). 
\subsection{Aside: Broken symmetries and residual symmetries}
\label{SEC:Aside} 
In order to avoid confusion, let us be quite clear about the pattern 
of the symmetry breaking emerging from our picture of the random 
solidification transition.  The liquid state, even if crosslinks 
are present, is fully translationally invariant.  Crosslinks do not,  
themselves, explicitly break translational symmetry, even though 
their presence in sufficient numbers can cause translational 
symmetry to break spontaneously, as occurs at the transition to 
the random solid state, in which particles become spontaneously 
localized in space.  All this applies to specific realizations 
of crosslinking.  

Now, what about the permutation symmetry of the replicated, 
disorder-averaged theory, i.e., what about replica symmetry?   Whether or
not this symmetry also spontaneously breaks at the  transition is only an
elaboration on top of the fundamental process,  namely translational
symmetry breaking.  To date, limited searches  have been undertaken (see
Refs.~\cite{REF:GZ:rsb,REF:PG:rsb}) but no stable saddle points having
broken replica symmetry have been found.  Moreover, local stability has
been established (see  Ref.~\cite{REF:CGZ:stability}) for the saddle point
described in Sec.~\ref{SEC:RSMRFT}.  This saddle point has the symmetries 
of the permutation of the replicas and the common translations and 
rotations of the replicas.  What have been lost at the transition are the 
symmetries of the relative translations and rotations of the replicas. 
\section{Emergent elasticity}
\label{SEC:Elastic} 
Perhaps the most remarkable (and certainly the most useful) feature 
of the phase of matter obtained via vulcanization is its elasticity.  
I will not dwell on this topic here; it is discussed in 
Refs.~\cite{REF:ENVT,REF:Rigidity}.  I will just mention  that, at 
least at the level of mean-field theory, one can address rather 
directly the following question: By how much does the free energy 
density increase when a shear (i.e.~volume-preserving) deformation 
is applied to sample of vulcanized matter?  In answering this question 
one finds that the random solid state is a homogeneous, isotropic 
elastic medium, characterized by a standard elastic free energy.  
Moreover, one finds that the shear modulus is (essentially) proportional 
to the temperature, confirming that the elasticity is primarily 
entropic in origin, and that this modulus vanishes as the third power 
of the excess constraint density, as the transition is approached from 
the liquid side.  Some of the probabilistic features of how random solids 
deform under shear (which seem rather counter-intuitive, at least to me) 
are also discussed in Refs.~\cite{REF:ENVT,REF:Rigidity}. 
\section{Long-range correlations and divergent susceptibilities 
for random solidification}
\label{SEC:DSRS} 
Let us now turn to the issue of long-range correlations that mark 
the onset of random solidification, and the attendant issue of 
diverging susceptibilities.  In the simpler context of, say, the 
ferromagnetic Ising transition, the two-point spin-spin correlation 
function quantifies the notion that the \lq\lq by hand\rq\rq\ alignment 
of one particular spin induces appreciable alignment of all the spins 
within roughly one correlation length of it, this distance growing 
as the transition is approached.  Now imagine approaching the random 
solidification transition from the liquid side.  The incipient order 
involves random localization and so, by analogy with the ferromagnetic 
case, the appropriate correlation function is the one that answers the 
question: Suppose that a particle is localized \lq\lq by hand\rq\rq; to 
what extent and over what spatial region would other particles respond 
by becoming localized?  

This scenario is, of course, the application to random 
solidification of the percolation theory question: What is the 
likelihood that two sites separated by a certain distance will be in 
the same cluster?  After all, if two macromolecular segments are 
connected to the same cluster then the \lq\lq by hand\rq\rq\
localization of one would cause the localization of the other.  Note, 
however, the entire physical domain that is absent from any percolation 
picture, namely, the thermal fluctuations of the particles:  
Not only do the relative separations of the particles on the cluster 
fluctuate but also the state in question is liquid, so that the particles 
are only relatively localized, not localized in space. 

Bearing these remarks in mind, let us consider the basic correlator 
associated with the $\Omega$ field theory:  
$\big\langle\Omega({\hat{k}_1})\,
             \Omega({\hat{k}_2})\big\rangle$.  
Not surprisingly, given that $\Omega$ is about to acquire a nonzero 
expectation value, this correlator becomes long-ranged at the transition.  
Indeed, at the Gaussian level one has 
\begin{equation}
\big\langle
\Omega({\hat{k}_1})\,
\Omega({\hat{k}_2})
\big\rangle\propto
\delta_{\hat{k}_{1}+\hat{k}_{2},\hat{0}}
\,\,\big/\,
\big(-2a\epsilon+b\hat{k}^{2}_{1}\big)\,.
\label{EQ:TPGauss}
\end{equation}
But what about a physical interpretation?  Well, as we have discussed 
in Sec.~\ref{SEC:CRSS}, the field $\Omega$ is closely related to the 
order parameter capable of detecting the localization associated with 
random solidification and, thus, the growing correlations of $\Omega$ 
should foretell incipient random localization, and they do.  To see 
this, consider the construct  
\begin{eqnarray}
C_{\bf q}({\bf r}-{\bf r}^{\prime})
\propto
\bigg[
{1\over{N}}\sum_{j,j^{\prime}=1}^{N}
\big\langle
\delta^{(d)}\big({\bf r}-{\bf R}_{j}\big)\, 
\delta^{(d)}\big({\bf r}^{\prime}-{\bf R}_{j^{\prime}}\big)\, 
\big\rangle
\nonumber\\
\lo
\qquad\qquad\qquad\quad
\times
\big\langle
\exp\big( i{\bf q}\cdot({\bf R}_{j}-{\bf r})\big)\, 
\exp\big(-i{\bf q}\cdot({\bf R}_{j^{\prime}}-{\bf r}^{\prime})\big) 
\big\rangle
\bigg]. 
\label{EQ:PhyDef}
\end{eqnarray}
For ${\bf q}={\bf 0}$ this is simply proportional to the density-density 
correlation function and, as such, is not of central importance at the 
random solidification transition.  However, for ${\bf q}\ne{\bf 0}$ it 
addresses the question:  If a particle near ${\bf r}$ is localized on 
the scale $q^{-1}$ (or more strongly), how likely is a particle near 
${\bf r}^{\prime}$ to be localized on the same scale (or more strongly)? 
It is straightforward to show that $C_{\bf q}(\brho)$ is related to the 
$\Omega$-$\Omega$ correlator: 
\begin{eqnarray}
\fl
C_{\bf q}(\brho)
\propto
\lim_{n\to 0}\,
\sum\nolimits_{\bf k}
\rme^{\rmi({\bf k}-{\bf q})\cdot\brho}\,
\big\langle
\Omega({\bf 0}, {\bf q},-{\bf k},{\bf 0},\ldots,{\bf 0})\,
\Omega({\bf 0},-{\bf q}, {\bf k},{\bf 0},\ldots,{\bf 0})
\big\rangle.
\label{EQ:connect}
\end{eqnarray}
The ${\bf q}\to{\bf 0}$ limit of  $C_{\bf q}(\brho)$
determines how likely it is for two particles a distance 
$\vert\brho\vert$ apart to be connected in a cluster, i.e., to be 
mutually localized, regardless of the strength of this localization. 
To construct the corresponding divergent susceptibility 
we integrate over space and pass to the ${\bf q}\to{\bf 0}$ 
limit, thus obtaining a measure of the spatial extent over which 
pairs of particles are mutually localized: 
\[
\fl
\lim_{{\bf q}\to{\bf 0}}
\int\rmd^{d}\rho\,
C_{\bf q}(\brho)
\propto
\lim_{{\bf q}\to{\bf 0}}
\lim_{n\to 0}\,
\left\langle
\Omega({\bf 0}, {\bf q},-{\bf q},{\bf 0},\ldots,{\bf 0})\,
\Omega({\bf 0},-{\bf q}, {\bf q},{\bf 0},\ldots,{\bf 0})
\right\rangle
\propto
\vert\epsilon\vert^{-\gamma}. 
\]
At the Gaussian level of approximation, this 
susceptibility diverges with the classical exponent $\gamma=1$. 
\section{Critical fluctuations}
\label{SEC:Critical} 
So far, our exploration of the random solidification transition has 
been at the level of mean-field theory (for the order parameter, 
elasticity and stability with respect to small fluctuations of the 
solid state), and the Gaussian approximation (for the correlations in 
the liquid state).  What about critical fluctuations?  In the present 
section I shall report some results on this issue that have been 
obtained very recently in collaboration with Weiqun 
Peng~\cite{REF:PG:VTBMFT} and primarily concern critical 
fluctuations in the liquid state and at the critical point. 
\subsection{Replica field theory}
\label{SEC:RFT} 
Let us progress beyond mean-field theory and the Gaussian approximation 
by regarding the Eq.~(\ref{EQ:LFE}) not as a Landau theory but as a 
Landau-Wilson effective Hamiltonian for a field-theoretic approach to 
the issue of critical fluctuations.  As such, it contains a Gaussian 
term and a cubic interaction term which, by na{\"\i}ve dimensional 
analysis, can be seen to be the most relevant relevant perturbation 
below six spatial dimensions (at least in the $n\to 0$ limit).  

It is worth noting the similarities and distinctions between this 
field theory and the (cubic, $n+1$-state Potts) field theory, the 
$n\to 0$ limit of which can be invoked to study 
percolation~\cite{REF:TCL:LH31}: 
\begin{equation}
\fl
\int_{V}\rmd^{d}r\,
        \Big(
\,\sum_{\alpha=1}^{n}
\left(
 {1\over{2}}t\,\psi_{\alpha}^{2}
+{1\over{2}}\vert\bnabla\psi_{\alpha}\vert^{2}
\right)
-w^{(3)}\sum_{\alpha,\beta,\gamma=1}^{n}
 \lambda^{(3)}_{\alpha\beta\gamma}\,
 \psi_{\alpha}\,
 \psi_{\beta}\,
 \psi_{\gamma}
        \Big),
\label{EQ:Potts}
\end{equation}
where $t$ controls the bond-occupation probability (and hence the
percolation transition), $w^{(3)}$ is the nonlinear coupling, and
$\lambda^{(3)}_{\alpha\beta\gamma}$ is the \lq\lq Potts tensor\rq\rq\
(which controls the internal symmetry of the theory).  As 
for similarities, there is the cubic nature of the interaction, the 
$(n+1)$-fold permutation symmetry, and the passage to the $n\to 0$ 
limit.  As for distinctions, the Potts field theory has a real 
multiplet of $n$ fields on $d$-dimensional space; the vulcanization 
field theory has a real field living on $(n+1)$-fold replicated 
$d$-dimensional space.  Furthermore, 
the Potts field theory represents a setting involving a {\it single\/} 
ensemble, the ensemble of percolation configurations, whereas the 
vulcanization field theory describes a physical problem in which  
{\it two\/} distinct ensembles (thermal and disorder) play essential 
roles.  As such, the latter is capable of providing a unified theory 
not only of the transition but also of the structure, correlations and 
(e.g.~elastic) response of the emerging random solid state.
\subsection{Recovering de~Gennes' Ginzburg criterion}
\label{SEC:DGGC} 
As we shall report in detail in Ref.~\cite{REF:PG:VTBMFT}, suppose 
we take the $\Omega$ field theory~(\ref{EQ:LFE}) and try to assess 
the range of constraint densities $\delta\epsilon$ (around the 
critical constraint density) within which fluctuation corrections 
are significant, i.e., to construct a Ginzburg criterion.  Then, 
provided we correctly account for the non-critical nature of the 
density fluctuations, we find that fluctuations increase the 
critical crosslink density (as one would expect on general grounds), 
along with a Ginzburg criterion which, for the case of vulcanized 
macromolecular matter, reads
\begin{equation}
\delta\epsilon
\approx 
\big(L/\ell\big)^{-(d-2)/(6-d)}\,\varphi^{-2/(6-d)}, 
\label{EQ:Ginzburg}
\end{equation}
where $L/\ell$ is the number of (essentially independent) segments 
on each chain and $\varphi$ is the volume-fraction taken up by the 
macromolecules: e.g., for $d=3$ shorter chains show stronger critical 
effects.  This dependence on $L/\ell$ is precisely that derived 
long ago by de~Gennes on the basis of a percolative 
picture~\cite{REF:deGennes}.  
\subsection{Expansion about six spatial dimensions}
\label{SEC:EASD} 
As we shall also report in detail in Ref.~\cite{REF:PG:VTBMFT}, we 
have, in addition, implemented a momentum-shell renormalization-group 
approach the the field theory~(\ref{EQ:LFE}) in order to construct 
universal quantities in an expansion around six spatial dimensions. 
Here, too,  one has to exercise considerable care in handling the 
vital constraint on the fields associated with the suppression of 
density fluctuations by particle interactions.  What emerges, at 
least to first order in $6-d$ (and likely beyond), is the following 
striking result: the critical state is governed by flow equations 
isomorphic to those emerging from the percolation limit of the Potts 
field theory.  Thus, e.g., to first order in $6-d$ the critical 
exponents 
$\eta$ [$=-(6-d)/21$, 
which describes the decay of the order parameter fluctuation 
correlations at the vulcanization transition], 
$\nu^{-1}$ [$=2-(5(6-d)/21)$, 
which describes the divergence of the fluctuation correlation 
length], 
and $\beta$ [$=1-((6-d)/7)$, 
which describes the growth of the localized fraction]
take on precisely the values that one would expect by examining 
analogous quantities in percolation theory. 

Although it is tempting to ask: \lq\lq How could it be 
otherwise?\rq\rq, it must be borne in mind that the vulcanization 
field theory is not the same as the Potts field theory.  It is 
richer, as it emerges from an underlying semi-microscopic 
picture in which there are both quenched and annealed variables, 
and contains a detailed description of the structure of the 
emergent random solid state in addition to being capable of 
capturing the critical properties of the vulcanization transition.  

The nature of the vulcanization transition and its relationship 
with the percolation transition in and around two spatial 
dimensions is an intriguing issue, as ongoing work with 
H.~E.~Castillo and W.~Peng is revealing~\cite{REF:CGP:wip}.  
Things become especially fascinating in two dimensions in view of 
the fact that there is a conventional percolation transition in 
there, whereas strong evidence (in the form of Goldstone 
modes resulting from the spontaneously broken {\it continuous\/} 
translational symmetry) suggests that two is the lower critical 
dimension for the vulcanization transition.  Thus, it is tempting 
to speculate that the following scenario holds in two 
dimensions~\cite{REF:CGP:wip}:  
(i)~With a subcritical density of constraints, the network does 
not percolate, there is no infinite cluster, the vulcanization 
order parameter is zero, and its correlations decay exponentially 
with distance.  
By contrast, (ii)~with a supercritical density of constraints, 
the network percolates and there is an infinite cluster; but 
thermal fluctuations in the positions of the constituents overwhelm 
the tendency for true localization so that the amorphous 
solidification order parameter remains zero and, instead, its 
fluctuations decay algebraically with distance.  One might say that 
(constraint-density controlled) cluster fragmentation, rather than 
thermally-excited lattice defects, mediate the melting transition. 
If this scenario happens to be borne out, one would have a 
quasi-amorphous solid state---the random analogue of a two-dimensional 
solid~\cite{REF:DRN:DandG}---exhibiting quasi-long-range positional 
order but of a random rather than regular type. 
\section{Where next?}
\label{SEC:WNext} 
In addition to a full renormalization-group based exploration of 
amorphous solidification in and near two dimensions, touched upon 
in the previous paragraph, issues that I would very much like to 
see addressed further include:
(i)~dynamics near the vulcanization transition, including the 
divergence of the viscosity as the transition is approached from 
the liquid side;
(ii)~a full renormalization-group approach to the random solid 
state itself; and 
(iii)~quantized networks, perhaps a somewhat academic topic, but 
I think it would be fascinating to have at hand a model showing 
that rigidity (perhaps one might call it Casimir rigidity) can be 
acquired by random solids via {\it quantum\/} fluctuations, rather 
than the thermal fluctuations discussed here.  
I would also very much welcome further experiments, such as 
quasi-inelastic neutron scattering, aimed, e.g., at extracting 
information about the distribution of localization lengths near 
the random solidification transition. 
\ack 
I wish to thank the conference organizers for their vision and effort, 
which led to such a stimulating workshop.  I also wish to acknowledge 
support by the U.S.~National Science Foundation via grant DMR99-75187, 
as well as the contributions of several collaborators, most notably  
N~D~Goldenfeld, A~Zippelius, H~E~Castillo, W~Peng and K~A~Shakhnovich, 
and many useful discussions with A~Halperin. 

\end{document}